\documentclass[preprint,prl]{revtex4}

\usepackage{graphicx}
\usepackage{dcolumn}
\usepackage{amsmath}
\usepackage{bm}
\begin{document}
\title {Comment on ``Imaging the Local Density of States of Optical Corrals''}
\author{S. Savasta, O. Di Stefano, and R. Girlanda}
\affiliation{INFM and Dipartimento di Fisica della Materia e
Tecnologie Fisiche Avanzate, Universit\`{a} di Messina Salita
Sperone 31, I-98166 Messina, Italy}
\author{M. Pieruccini}
\affiliation{CNR, Istituto per i Processi Chimico-Fisici Sez.\ Messina,Via La Farina 237, I-98123 Messina, Italy}


\maketitle

{\bf{\center Comment on ``Imaging the Local Density of States of Optical Corrals'' }}\\

In a recent letter Chicanne {\em et al.} [1] reported the experimental observation of the electromagnetic local density of states LDOS established by gold nanostructures. The obtained images have been compared
with combinations of partial  LDOSs defined in terms of the imaginary part of the Green-tensor ${\bf G}^I = [{\bf G}-{\bf G}^\dag]/(2i)$ calculated at the tip position. Moreover just this comparison was the criterion for the choice  of the optimum tip design. These results support the point of view that
${\cal G}_{\bf u} =-({2 \omega}/{\pi c^2} )\, {\bf u} \cdot {\bf G}^I({\bf r}, {\bf r}, \omega) \cdot {\bf u}$ (${\bf u}$ is the unit vector used to define the effective dipole associated to the illuminating tip) is the key quantity to interpret SNOM images in analogy with the electronic LDOS measured by the scanning tunneling microscope (STM).
Rigorous Green-tensor analysis shows that ${\cal G}_{\bf u}$ (that is also the key quantity determining spontaneous decay rates of molecular transitions) is not the correct key quantity, and that measurements in Ref.\ [1] should have been compared with a different quantity. Moreover the identification of  ${\cal G}_{\bf u}$ with the detected SNOM signal can lead to unphysical results.

Let us consider a device (the ideal SNOM in the illumination mode) whose illuminating tip can be modeled by a point-like source in the near-field region of a sample. The device should also be able  to detect the  scattered light over $4 \pi$ steradians in the far-field region. On the basis of the reciprocity theorem that allows the exchange of sources and detectors [2], the obtained signal is proportional to the local density of scattering modes  i.e.
the light intensity at a point ${\bf r}$  when the
material system is illuminated isotropically and incoherently  (e.g by a thermal source [3]). For a  point detector polarized along a specific direction ${\bf u}$, it can be expressed as
${\cal S}_{\bf u} = {\bf u} \cdot  {\bf S}({\bf r}, {\bf r}, \omega) \cdot {\bf u}$, where
${S}_{ij}({\bf r}, {\bf r}', \omega)= \sum_{\alpha}
 {\boldsymbol \psi}_{\alpha i}({\bf r}, \omega) {\boldsymbol \psi}^*_{\alpha j}({\bf r}', \omega)$, with
${\boldsymbol \psi}_{\alpha} = {\boldsymbol \Omega}{\boldsymbol \psi}^0_{\alpha}$
describing the electric field in presence of the scattering system arising from an input plane wave ${\boldsymbol \psi}^0_{\alpha}(\omega)$  with given incident direction and polarization [$\alpha = (\hat{\bf k},p=1,2 )$]. The M\o ller operator ${\boldsymbol \Omega}$ is given by ${\boldsymbol \Omega}= \left[ {\bf 1} -{\bf G}(\omega){\bf e}_s(\omega) \right]$, where  ${\bf e}_s \equiv k^2 \chi_{ij}({\bf r},\omega)$ is related to the susceptibility tensor of the scattering system.
Using the  relation ${\boldsymbol \psi}_{\alpha} = {\boldsymbol \Omega}{\boldsymbol \psi}^0_{\alpha}$,
ones finds ${\bf S}= {\boldsymbol \Omega}{\bf S}^0{\boldsymbol \Omega}^\dag$.
Introducing the well-known relationship $
{\bf S}^0 = (-2 \omega/\pi c^2){\bf G}^{0I}$ for free space, and using the Dyson equation
(${\bf G} = {\boldsymbol \Omega}{\bf G}^0$), it is easy to obtain:
${\bf S} =   -({2 \omega}/{\pi c^2} ){\bf G}^I - {\bf A}$, with
${\bf A} = ({2 \omega}/{\pi c^2} ) {\bf G}({\bf e}^I_s ){\bf G}^\dag $.
As a consequence what is measured by an ideal SNOM set up in illumination mode ${\cal S}_{\bf u}$, differs from ${\cal G}_{\bf u}$. The difference  between
these two quantities (${\cal A}_{\bf u} = {\bf u} \cdot {\bf A}({\bf r}, {\bf r},\omega) \cdot {\bf u}$) plays  a key role in near-field thermal emission [3,4].
In presence of lossy media as metallic nanostructures ($\chi^I \neq 0$),  ${\cal A}_{\bf u}$  can be of the same order of magnitude of
${\cal G}_{\bf u}$, thus  ${\cal G}_{\bf u}$ and ${\cal S}_{\bf u}$ can present qualitative as well as quantitative differences as shown in Fig.\ 1.
\begin{figure}
  [h]\centerline{\scalebox{1.}{\includegraphics{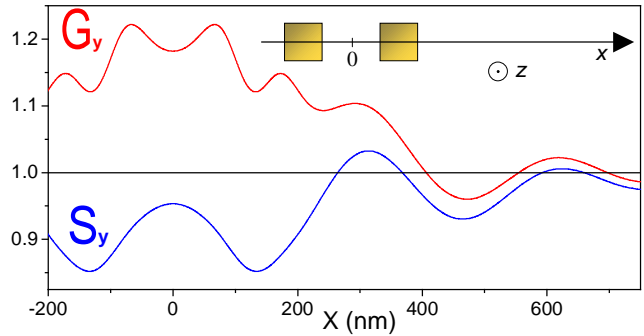}}}
 \caption{
 (Color online) Cross-cuts (see inset) of ${\cal G}_{\bf y}$  and ${\cal S}_{\bf y}$ 
 calculated (by the Green-tensor technique) at constant height ($70$ nm) above two gold pads of area $100 \times 100$ nm$^2$ and height $50$ nm lying on an ITO substrate.
}\label{fig1}
  \end{figure}
Let us consider the case of a system supporting bounded modes (like e.g. the whispering gallery modes of a dielectric sphere). Although these modes can be excited by a point-like source close to the system,
they do not contribute to the detected far-field signal, owing to their evanescent character.
This behavior is well reproduced by ${\cal S}_{\bf u}$ that does not contain contributions from bound modes
(${\boldsymbol \psi}_{\alpha}$ are by definition scattering modes).
On the contrary ${\cal G}_{\bf u}$ includes these resonances and clearly leads to unphysical results if it is interpreted as the SNOM signal. Also we find by scattering calculations that in presence of media that display gain in some frequency range ($\chi^I \leq 0$), ${\cal G}_{\bf u}$ can be negative (in contrast with ${\cal S}_{\bf u}$)
thus failing in describing a power signal measured by a photodetector.

In conclusion, we show that the correct key quantity  for the interpretation of SNOM images acquired in the illumination mode
is different from ${\cal G}_{\bf u}$  also for the ideal illumination-mode  setup (point-like source, $4 \pi$ steradiands far-field detection). On the contrary the density of scattering modes ${\cal S}_{\bf u}({\bf r}_{tip},\omega)$ describes the functionality of the ideal illumination-mode  setup and can be used to interpret non-ideal SNOM images in analogy with the electronic LDOS used to interpret STM images.
Finally it is worth noting that for systems with negligible absorption (not the case of Ref.\ [1]), and not supporting bound modes, ${\cal S}_{\bf u}({\bf r}, \omega)= {\cal G}_{\bf u}({\bf r}, \omega)$.

\newpage

\section{List of Figure Captions}
 \vspace{1cm} \noindent
Fig.\ 1  (Color online) Cross-cuts (see inset) of ${\cal G}_{\bf y}$  and ${\cal S}_{\bf y}$  calculated (by the Green-tensor technique) at constant height ($70$ nm) above two gold pads of area $100 \times 100$ nm$^2$ and height $50$ nm lying on an ITO substrate.
\newline


\begin{thebibliography}{50}

\bibitem{LDOS3} C.\ Chicanne {\em et al.}, Phys. Rev.\ Lett.\
{\bf 88}, 097402 (2002).


\bibitem{Mendez} E.\ Mendez {\em et al.}, Opt.\ Commun.\ {\bf 142}, 7 (1997).



\bibitem{prasavasta} S.\ Savasta {\em et al.},
Phys. Rev. A {\bf 65}, 043801 (2002).



\bibitem{greffet2} A.\ V.\ Shchegrov {\em et al.}, Phys.\ Rev.\ Lett.\ , {\bf 85},
1548 (2000).




\end{thebibliography}
\end{document}